\documentclass[10pt, a4paper]{article}

\usepackage{indentfirst}
\usepackage{misccorr}
\usepackage{graphicx}
\usepackage{amsmath}
\usepackage{pstool}
\usepackage{hyperref}
\usepackage{amssymb}
\usepackage{tocbibind}
\usepackage[normalem]{ulem}
\usepackage{authblk}

\makeatletter
\renewcommand{\appendix}
{
	\par
	\setcounter{section}{0}
	
}

\makeatother

\hypersetup{unicode=true}

\begin{document}

	\title{Transition between metastable equilibria: applications to binary-choice games}
	\author{A.~Antonov$^{(1)}$\footnote{antonov@lpi.ru } }
	\author{A.~Leonidov$^{(1,2)}$}
	\author{A.~Semenov$^{(1,3,4)}$}
	\affil{{\small (1) P.N. Lebedev Physical Institute, Moscow, Russia\\
			(2) Moscow Institute of Physics and Technology, Dolgoprudny, Russia\\
			(3) Skolkovo Institute of Science and Technology, Moscow, Russia\\
			(4) HSE University, Moscow, Russia}}
	\date{}
	
	\maketitle
	
	\begin{abstract}
		
	 Transitions between metastable equilibria in the low-temperature phase of dynamical Ising game with activity
	 spillover are studied in the infinite time limit. It is shown that exponential enhancement due to activity spillover,
	 which takes place in finite-time transitions, is absent in the infinite time limit. In order to demonstrate that, the
	 analytical description for infinite time trajectory is developed. An analytical approach to estimate the probability
	 of transition between metastable equilibria in the infinite time limit is introduced and its results are compared
	 with those of kinetic Monte Carlo simulation. Our study sheds light on the dynamics of the Ising game and has
	 implications for the understanding of transitions between metastable states in complex systems. 
		
		\medskip
		
	\end{abstract}
	

	\maketitle
	
	\newpage
	
	\section{Introduction}
	
	Studies of noisy binary choice games are of special interest because of the existence of close parallels to statistical physics of spin systems, in particular to static and dynamic properties of phase transitions in them \cite{Blume2003,Bouchaud2013,Salinas2001}. These parallels are particularly intriguing because of the fundamentally different  origins of equilibria in game theory and statistical physics: in game theory equilibration is a result of balancing individual interests while in statistical physics equilibration is a search of a global minimum of free energy. For the noisy binary choice problem on complete graphs it is long known (see \cite{Blume2003} and references therein), that for a special choice of noise game-theoretic equilibria are characterized by the same mean-field Curie-Weiss equation as that describing phase transitions in magnetics; see, e.g.,\ \cite{Salinas2001}. The properties of static and dynamic equilibria in noisy binary choice games were studied in \cite{Leonidov2019,leonidova2020qre,leonidov2021ising} for arbitrary noise and complete and random graph topologies. It was established in particular that static game-theoretic equilibria in noisy binary choice games on graphs correspond to the so-called quantal response or expectation equilibria \cite{Goeree2016}.
	
	The dynamics of games can, however, be fundamentally different from conventional spin dynamics due to a variety of possible mechanisms. One of these is a possibility of activity spillover (self-excitation) that was intensively studied for so-called Hawkes processes \cite{Hawkes1971} with applications to finance \cite{Filimonov2015,Hardiman2013}, earthquakes \cite{Ogata1988}, and other subjects, see the recent review in \cite{Laub2015}. A master equation formalism for such processes was developed in \cite{Kanazawa2020_1,Kanazawa2020_2}.
	The effects of an activity spillover different from the Hawkes self-excitation mechanism for a noisy binary choice game (Ising game) on complete graphs was studied in \cite{Antonov2021}. The main focus of \cite{Antonov2021} was in studying transitions between metastable equilibria in the low-temperature phase taking place at finite time. It was observed that activity spillover leads to an exponential acceleration of such transitions.
	The present paper complements the analysis of \cite{Antonov2021} by studying transitions between metastable equilibria in the limit of infinite time. The importance of studying this limit is, first, in establishing a link with a rich literature on Kramers rate \cite{Kramers1940} and, second, in that in this limit the exponential enhancement is absent and an analysis of pre-exponential contribution is necessary. In analyzing this problem we develop an analytical description of the infinite-limit trajectory and suggest an analytical formula for the transition rate that is compared with the results of exact numerical simulations.  
	
	\section{Model}
	\label{sec:2}
	
	We consider an Ising game \cite{Leonidov2019,leonidova2020qre,leonidov2021ising}, i.e.\ a dynamical noisy binary choice game of $N$ agents on a complete graph topology. Each agent $i$ has two possible strategies $s_i=\pm1$ so the system is fully described by the vector ${\bf s}_t = (s_1, \ldots , s_n)_t$ at given time $t$. The temporal evolution of the strategies configuration ${\bf s}_t \to {\bf s}_{t+\delta t}$ within a small time interval $\delta t $ is assumed to be driven by a strategy flip  $s_i \to -s_i$ of some agent $i$ with the flip probability
	\begin{eqnarray}
		\textrm{Prob}[s_i \to -s_i|(t;t+\delta t)] = \lambda (t)  \delta t \; \gamma_i (s_i \to -s_i \vert {\bf s}_{-i,t}) \nonumber \\
		\;
	\end{eqnarray}
	where $\lambda(t) $ is an activity rate that in the considered case of complete graph topology is the same for all agents, so that  $\lambda (t) \delta t$ is a time-dependent probability for some agent $i$ to be active and have a possibility to change a strategy within a time interval $(t,t+\delta t)$ while $\gamma (s_i \to -s_i \vert {\bf s}_{-i,t})$ is a probability, for an active agent $i$, of a strategy flip dependent of the current configuration ${\bf s}_{-i,t}$ of strategies in the closed neighborhood of this node \footnote{In computer science and graph theory, the closed neighborhood consists only of the nodes adjacent to the given node, excluding the node itself.}. In what follows we shall assume a noisy best response (Ising-Glauber) flip rate \footnote{This choice corresponds to the Gumbel noise in the individual agents utilities.}. For a complete graph topology at large $N$, it is the same for all agents
	\begin{eqnarray}
		\gamma (m(t)) & = & \frac{1}{2}\left[1-s_i\tanh\left(\beta J m (t) \right)\right] \nonumber\\ \rightarrow \;\; \gamma_{\pm} (m(t)) & = & \frac{1}{2}\left[1 \pm \tanh(\beta J m(t))\right]
		\label{eq:defgamma}
	\end{eqnarray}
	where $\beta=1/T$ is an inverse temperature, $J$ is an Ising coupling constant, $\gamma_{\pm} = \gamma(\mp s \to \pm s)$ and $m(t) = \frac{1}{N} \sum_{i=1}^N s_i$.

	Our study is focused on manifestations of a specific excitation mechanism converting past strategy flips of all agents into an increase of the overall activity rate $\lambda$ introduced in \cite{Antonov2021}. An activity followed by a strategy flip can naturally be termed a realized activity. The excitation mechanism in which the realized activity amplifies the overall activity rate $\lambda$ can then be termed a realized activity spillover. The term spillover is widely used in theoretical economics and game theory for describing effects of propagation of certain properties or signals onto neighboring agents; see, e.g.,\ \cite{sanna2007multinational,nie2022technology}.
	
	A quantitative description of the activity spillover can naturally be given in terms of a general Hawkes process
	\begin{equation}\label{eq:hawkes0}
		\lambda(t)=\lambda_0 + \mu \sum_{\tau_k < t} \phi (t-\tau_k) 
	\end{equation}
	where $\{ \tau_k\}$ are times at which strategy flip of one of agents took place, $\mu \geq 0$ quantifies the spillover strength and and $\phi(t-\tau_k)$ is a memory kernel. In the absence of long-memory effects a natural choice for the memory kernel in \eqref{eq:hawkes0} made in the original paper \cite{Hawkes1971} and many followup ones (see, e.g.,\ \cite{Laub2015,Filimonov2015,Kanazawa2020_1,Kanazawa2020_2}) is an exponential function. With this choice \eqref{eq:hawkes0} takes the form
	\begin{equation}\label{eq:hawkes}
		\lambda(t)=\lambda_0 +\frac{\mu}{N} \sum_{\tau_k < t}e^{-b(t-\tau_k)},
	\end{equation}
	where $b$ parametrizes memory depth and the factor $1/N$ is a convenient choice for the considered case of complete graph topology.

	The case $\mu=0$ corresponds to a standard Poisson dynamics of the system with constant (Poisson) intensity $\lambda(t) = \lambda_0$ considered in, e.g.,\ \cite{Blume2003,Bouchaud2013,Leonidov2019,leonidov2021ising}. In what follows we shall term the corresponding game a Poisson Ising game. In the case $\mu>0$ with the realized activity spillover switched on, the corresponding game will be termed the Hawkes Ising game.
	
	In the Hawkes Ising game a state of the system is fully specified by $m(t)$ and $\lambda(t)$. Its dynamics is described by the probability distribution $P(m,\lambda;t)$ and is specified by the  parameters 
	$\lambda_0,\, \mu,\, b,\, \beta J$. The particular case of $\mu = 0$ corresponds to a standard
	Poisson dynamics of the system with constant (Poisson) intensity
	$\lambda(t) = \lambda_0$ so that the probability distribution describing
	it is reduced to $P(m(t); t)$. This special case will in what follows be used as a reference model.
	
	In the limit $N \to \infty$, the probability density function $P(m,\lambda;t)$ can be described by the approximate Fokker-Planck equation derived in \cite{Antonov2021}
	\begin{eqnarray}
		\partial_t P & = & \partial_i (f_i P)+\frac{1}{N}\partial_i\partial_j \left(g_{ij} P\right) \\
		f_i & = & \begin{pmatrix}\lambda\left[ m-\tanh(\beta J m) \right] \\  -\lambda \left[ 1-m\tanh(\beta J m) \right]  +b \left[ \lambda-\lambda_0 \right] \end{pmatrix} \nonumber \\
		g_{ij} & = & \begin{pmatrix} \lambda \left[ 1-m\tanh(\beta J m) \right] & -\lambda\left[ m-\tanh(\beta J m) \right] \\  -\lambda\left[ m-\tanh(\beta J m) \right] & \lambda \left[ 1-m\tanh(\beta J m) \right] \end{pmatrix},\nonumber
		\label{eq:FPE}
	\end{eqnarray}
	where summation over repeated indices is assumed. Here and in what follows the indices $i$ and $j$ represent coordinates $m$ and $\lambda$, and the following rescaling was performed: 
	\begin{equation}
		\lambda \to \frac{2\lambda}{\mu}, \lambda_0 \to \frac{2\lambda_0}{\mu}, b\to \frac{2b}{\mu}, t \to \frac{\mu t}{2}.
		\label{eq:rescalation1}
	\end{equation}
	
	The Fokker-Planck equation \eqref{eq:FPE} describes Brownian motion in an external vector field $f_i$ in the plane $(\lambda, m)$ subject to noise effects described by the matrix $g_{ij}$ and corresponds to a mean field game-type description of the dynamic Ising game under consideration \footnote{The standard description of a mean field game includes, in addition to a Fokker-Planck equation, additional equations describing optimal control, see e.g.\ \cite{lasry2007mean}.}. Let us stress that evolution equation \eqref{eq:FPE} that is foundational for our study holds only for the considered case of exponential memory kernel in \eqref{eq:hawkes0}, i.e.,\ for the standard Hawkes process of the form \eqref{eq:hawkes0}. We also note that the considered system is symmetric with respect to $m$-axis, and the external field is non-gradient, i.e.,\ $\partial_{\lambda} f_m \ne \partial_m f_{\lambda}$.

	In such a parametrization, the process of realized activity spillover is controlled by a single memory kernel parameter $b$. In the special case of Poisson Ising game with $\mu = 0$ the rescaling in \eqref{eq:rescalation1} is of course not relevant and the corresponding one-dimensional dynamics along the $m$ axis can be studied by simply taking $\lambda(t) \equiv \lambda_0$.
	
	Depending on parameters of the system, the considered system can have different equilibrium configurations $(\lambda_{\rm{eq}}, m_{\rm{eq}})$ given by the  zeros of vector field $f_i = 0$ corresponding to stable fixed points. 
	As we can see from Eq.~\eqref{eq:FPE}, for both Hawkes and Poisson Ising games, for any time-dependent activity rate in the Hawkes Ising game $\lambda(t)$, the $m$-equilibria are described \cite{Antonov2021} by the same Curie-Weiss equation as in \cite{Blume2003,Bouchaud2013,Leonidov2019} 
	\begin{equation}
		m_{\rm{eq}} = \tanh(\beta J m_{\rm{eq}}).
	\end{equation}
	
	For high temperatures $\beta J < 1$ the system has one equilibrium $m_{\rm{eq}} = 0$, and for low temperatures $\beta J > 1$ it has two symmetrical (meta)stable equilibria at $m_{\rm{eq}} = \pm m_0(\beta)$ as well as the unstable one at $m=0$ serving as a separatrix separating the two stable ones.
	
	The $\lambda$ - equilibria are more complicated and depend on both temperature $\beta J$ and self-excitation memory kernel parameter $b$. 
	
	In the high-temperature phase $\beta J < 1$ and $b>1$, we have the equilibrium configuration of the form 
	$$m=0, \lambda = \frac{b\lambda_0}{b - 1}$$ 
	while for $b<1$ we have a blow-up solution with $\lambda \to \infty$ for $m = 0$.
	
	In the low-temperature phase $\beta J > 1$, the three following modes are possible:
	
	\begin{itemize}
		\item Mode 1 ``calm agents'': if $b > 1$, then we have two (meta)stable equilibrium configurations at $$m = \pm m_0(\beta), \lambda = \frac{b\lambda_0}{b - 1 + m_0^2(\beta)} = \tilde{\lambda}(m_0) $$ as well as the unstable saddle one at $$m = 0, \lambda = \frac{b\lambda_0}{b - 1}$$
		\item Mode 2 ``excited agents'': if $1 - m_0^2(\beta) < b < 1$, then we still have equilibrium configurations at $$m = \pm m_0(\beta), \lambda = \tilde{\lambda}(m_0),$$ but the saddle configuration is now absent: $$m = 0, \lambda \to \infty$$
		\item Mode 3 ``chaotic agents'': if $b < 1 - m_0^2(\beta)$, then $$\lambda \to \infty$$ for all extrema of the $m$-axis.
	\end{itemize}
	
\begin{figure}[htp!]
		\center{\includegraphics[width=\linewidth]{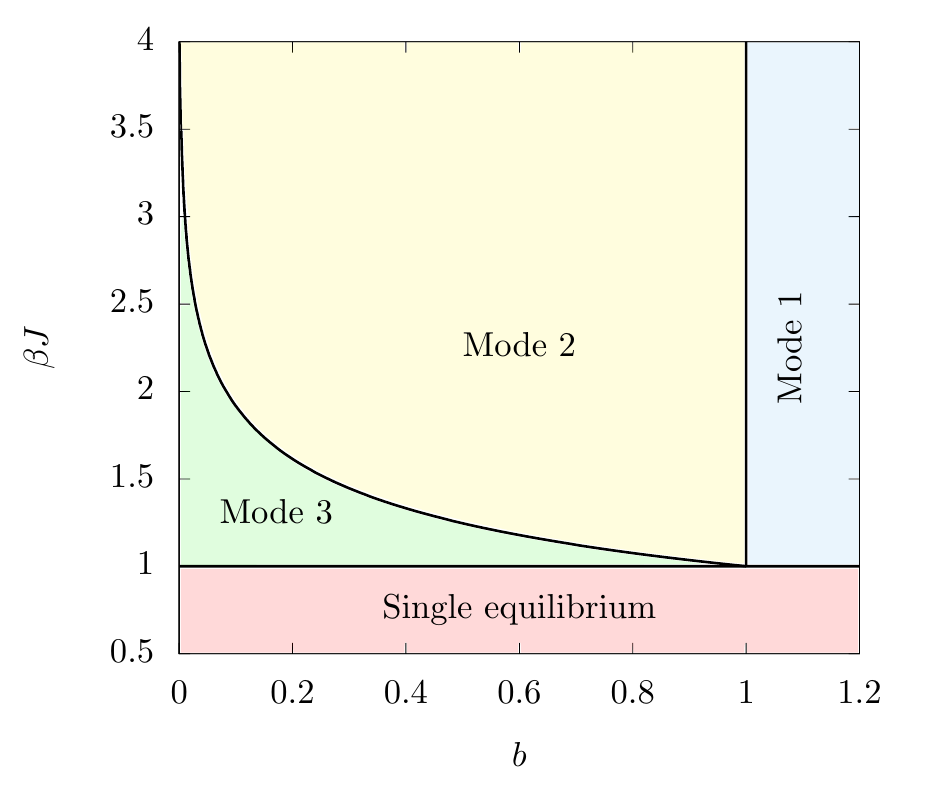}}
		\caption{Phase diagram of all possible modes in the $(b, \beta J)$ plane for $\lambda_0 = 1$. The red area (single equilibrium) here denotes the presence of only equilibrium along the $m$-axis. Blue, yellow and green areas correspond to Modes 1, 2 and 3, respectively (see the description in the main text).}
		\label{fig:phase}
	\end{figure}

	The phase diagram showing the above modes is given in Fig.~\ref{fig:phase}.
	
	At the timescale of $\tau_{\lambda} \sim 1/b$, in Modes 1 and 2 the system relaxes to the appropriate temperature-dependent equilibrium while the Mode 3 does not correspond to any equilibrium.  The dependence of such a relaxation on temperature $\beta J$ and Hawkes parameter $b$ in the Modes 1 and 2 was studied in \cite{Antonov2021}.
	
	At low temperatures $\beta J > 1$, the equilibrium configurations for Modes 1 and 2 are in fact metastable due to noise-induced transitions of the type $m_0(\beta) \leftrightarrow -m_0(\beta)$ taking place at large timescale $\tau \gg \tau_{\lambda}$. The saddle we introduced for Mode 1 then has the following physical meaning: it is the point where the transition trajectory from one equilibrium to another at the infinite time limit crosses the separatrix $m=0$. \cite{Feng2014}
	
	To consider these transitions, here and in what follows we fix $\beta J = 1.5$ to establish the mode with two metastable equilibria (Mode 1 or 2, see Fig.~\ref{fig:phase}). For our convenience, in what follows we shall consider the transition $-m_0(\beta) \rightarrow m_0(\beta)$.
	
	\section{Transition between metastable equilibria}
	
	\subsection{Long-time behavior of probability density function}
	\label{sec:3.1}
	
	The subject of our study is a comparison of the transition probability between the states $(m(t_a),\lambda(t_a))$ and $(m(t_b),\lambda(t_b))$ within the time interval $[t_a, t_b]$ for Hawkes and Poisson Ising games. In what follows we shall use a condensed notation $x_{a,b} = (m(t_{a,b}),\lambda(t_{a,b}))$ and fix $[t_a,t_b] = [0,\tau]$ so that the transition probability between two metastable states is
	
	\begin{equation}
		\mathcal{P}(x_b,t|x_a,0) \equiv P(x_b, t)\Big|_{x(0) = x_a}
		\label{eq:transition}
	\end{equation}
	where $m(0) = -m_0(\beta), \; \lambda(0) = \tilde{\lambda}(m_0)$ and $m(\tau) = m_0(\beta), \; \lambda(\tau) = \tilde{\lambda}(m_0)$. The transition probability \eqref{eq:transition} obeys \cite{Antonov2021} the Fokker-Planck equation \eqref{eq:FPE}.
	
	In the previous paper \cite{Antonov2021} we have compared the probabilities of transition between metastable equilibria in Hawkes and Poisson Ising games within a finite time interval $[0,\tau]$ and demonstrated an exponential acceleration of this transition in the Hawkes case. 
	These probabilities themselves, though, are exponentially small. The main goal of the present paper is to calculate this transition probability in the limit $\tau \to \infty$. To discuss this limit let us use, following \cite{Antonov2021}, the analogy 
	with classical mechanics. A formal justification for it can be found, e.g., in \cite{Maslov1981}.
	
	As the diffusion coefficient in \eqref{eq:FPE} is proportional to $1/N$, in the limit of $N \to \infty$, for solving the Planck equation we can use the WKB (Wentzel-Kramers-Brillouin) approximation. Introducing an analog of action $S(x,t)$ through $ P(x,t) \propto e^{-N S(x,t)} $,  we get the following Hamilton-Jacobi equation for $S$:
	\begin{eqnarray}
		\partial_t S(x,t) & = & f_i(x(t))\frac{\partial S(x,t)}{\partial x_i} \nonumber\\
		& - & g_{ij}(x(t))\frac{\partial S(x,t)}{\partial x_i}\frac{\partial S(x,t)}{\partial x_j}.
	\end{eqnarray}
	One can also introduce an analog of the Hamiltonian
	\begin{eqnarray}
		H(p,x;t) & = & - f_i(x(t))p_i(t) + g_{ij}(x(t))p_i(t)p_j(t), \\  p_i & = & \frac{\partial S}{\partial x_i}. \nonumber
	\end{eqnarray}
	The time evolution of the system is then given by the corresponding Hamilton equations
	\begin{eqnarray}
		\dot{x_i}(t) & = & - f_i(x(t)) + 2 g_{ij}(x(t)) p_j(t) \nonumber \\
		\dot{p_i}(t) & = & p_j\partial_i f_j(x(t)) -p_j \partial_i g_{jk}(x(t))p_k(t).
		\label{eq:EOM}
	\end{eqnarray}
	The system of Hamilton equations \eqref{eq:EOM} has the first integral $H(p,x) = E$. As will be shown later, the value of $E$ implicitly sets conditions on the transition time $\tau$ from one metastable equilibrium to another in the classical problem.
	
	The leading contribution to the transition probability has the form 
	
	\begin{equation}
		\mathcal{P}(x_i,x_f;\tau) \propto e^{-N\mathcal{S}}
	\end{equation}
	where the exponential factor $\mathcal{S}$ can be calculated by implementing the Maupertuis principle \cite{Landau}
	
	\begin{eqnarray}
		\mathcal{S} = S_0 - E\tau & = & \sum_{i}\int_0^{\infty} p_i(t) \dot{x}_i(t) dt - E\tau \nonumber \\
		& = & \sum_{i} \int\limits_{\rm{trajectory}} p_i dx_i - E\tau.
		\label{eq:action}
	\end{eqnarray}
	
	The abbreviated action according to this principle is stationary on the transition trajectory. The transition trajectory itself is determined by Eqs.~\eqref{eq:EOM}, the first integral $H(p,x) = E$ and, obviously, should minimize the trajectory-depended term $S_0$. Transition time is set by $E$ via relation $\tau=\partial S_0/\partial E$. \cite{Landau}
	
	In \cite{Antonov2021} we considered transition \textit{probability} from one metastable equilibrium to another in finite time ($E \ne 0$) and found out that the probability exponentially increases due to activity spillover. In the present study we augment the results of \cite{Antonov2021} by considering  introducing transition \textit{rates} in the infinite time limit corresponding to $E = 0$.
	
	The system of differential equation \eqref{eq:EOM} for $E = 0$ is solvable in quadratures. The corresponding solution for the transition trajectory can naturally be broken into two pieces.
	
	The first piece corresponding to transition from the initial equilibrium to separatrix $-m_0(\beta) \rightarrow 0$. The corresponding formulas read
	\begin{eqnarray}
		\dot{m}(t) & = & \lambda(t)[m - \tanh(\beta J m)], \label{solm}\\
		\dot{\lambda}(t) & = & \lambda(t)[1 - m\tanh(\beta J m)] - b(\lambda(t) - \lambda_0) \nonumber \\
		& - & \frac{\lambda(t)[m - \tanh(\beta J m)]^2}{1 - m\tanh(\beta J m)}, \\
		p_m & = & \frac{m - \tanh(\beta J m)}{1 - m\tanh(\beta J m)}, \\
		p_{\lambda} & = & 0. \label{wl}
	\end{eqnarray}
	The second piece corresponding to transition from the separatrix to another equilibrium $0 \rightarrow m_0(\beta)$ . The corresponding formulas read
	\begin{eqnarray}
		\dot{m}(t) & = & -\lambda(t)[m - \tanh(\beta J m)], \label{solg}\\
		\dot{\lambda}(t) & = & \lambda(t)[1 - m\tanh(\beta J m)] - b(\lambda(t) - \lambda_0), \\
		p_m & = & 0, \\
		p_{\lambda} & = & 0. \label{wgl}
	\end{eqnarray}
	We note that despite the symmetry with respect to $m$-axis, the transition trajectory is asymmetric as the external field is non-gradient. Eqs.~\eqref{solg}-\eqref{wgl} describe the classical equations of motion when descending from the saddle in the zero-noise limit. Eqs.~\eqref{solm}-\eqref{wl} are less trivial to obtain, but due to $m \to -m$ symmetry we can assume that Eq.~\eqref{solm} is analogous to Eq.~\eqref{solg} \cite{Kamenev2011}, and another trivial assumption $p_{\lambda} = 0$ provides us with the correct solution.
	
\begin{figure}[htp!]
		\center{\includegraphics[width=\linewidth]{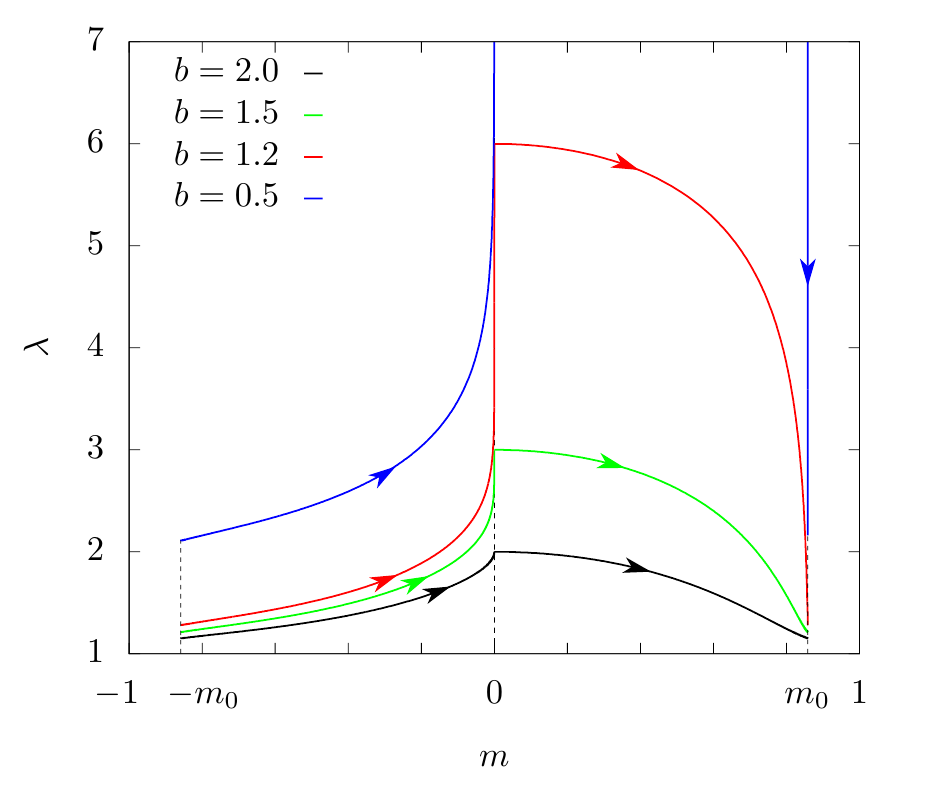}}
		\caption{Transition trajectories $\ -m_0(\beta) \to m_0(\beta)$ at the infinite time limit $E = 0$ given by Eqs.~\eqref{solm}-\eqref{wl} (left half) and Eqs.~\eqref{solg}-\eqref{wgl} (right half) for $b = 0.5,\, 1.2, \, 1.5, \, 2.0$ (trajectories with smaller $b$ have larger $\lambda$ values at the same $m$) at $\beta J = 1.5$, $\lambda_0 = 1$. The trajectories for Mode 1 ($b = 1.2, \, 1.5, \, 2.0$) are defined and have a discontinuity at the saddle, and the trajectory for Mode 2 ($b = 0.5$) diverges at $m = 0$.}
		\label{fig:traj}
	\end{figure}

	In accordance with the classification of modes introduced in Section~\ref{sec:2}, for different values of the parameter $b$ the Hawkes transition trajectory does either pass through the saddle point where it has the discontinuity (Mode 1) or diverges at the separatrix $m=0$ (Mode 2). The trajectories for various values of parameter $b$ are shown in Fig.~\ref{fig:traj}.
	
	From Eqs.~\eqref{eq:action},~\eqref{solm}-\eqref{wgl} it follows that in the infinite time limit for which $E = 0$, the exponential factor $\mathcal{S}$ is equal for Poisson and Hawkes Ising games for all $b$:
	
	\begin{equation}
		\displaystyle \mathcal{S} \hspace{1ex} = \int\limits_{-m_0(\beta)}^0\frac{m - \tanh(\beta J m)}{1 - m\tanh(\beta J m)} dm
	\end{equation}
	
	Therefore, for understanding a possible difference between the Hawkes and Poisson Ising games in the infinite time limit, an analysis of pre-exponential factor of the transition rate is required.
	
	\subsection{Pre-exponential factor of the transition rate}
	
	The calculation of the pre-exponential factor for the one-dimensional Poisson Ising game closely follows the original calculation by Kramers \cite{Kramers1940} and can be done analytically, see e.g.\ \cite{Caroli1981,Coleman1985}.  A more general result for larger number of dimensions, including the case of non-potential fields, was obtained in \cite{Bouchet2016}. However, this result is not applicable in our the two-dimensional Hawkes Ising game, since the transition trajectory in the non-gradient field has a discontinuity, see a related discussion in \cite{Cameron2018}.
	
	When the trajectory is defined (Mode 1), we can use analogies with one-dimensional motion. In the Kramers' problem for the potential with smooth barrier the pre-exponential factor of escape rate depends on second derivatives of the potential both for stationary attractor and a saddle.
	However, if the potential barrier is edge-shaped, then the result depends only on the second derivative of the potential at stationary attractor \cite{Matkowsky1982}.
	That leads us to an assumption that in the Hawkes Ising game acceleration with respect to Poisson Ising game is caused only by a corresponding change in the activity of agents in the equilibrium state, with the rest of motion having non-significant effect on the transition time.
	
	This assumption means that average transition times in the Hawkes Ising game and in the Poisson Ising game with intensity
	$\tilde{\lambda}(m_0)$ are equal. Therefore a ratio of transition
	times in the original Hawkes and Poisson Ising games can be
	written in the following form:
	
	\begin{equation}
		\frac{t_{\tiny{\textrm{tr,P}}}}{t_{\tiny{\textrm{tr,H}}}} \simeq \frac{\lambda_0}{\tilde{\lambda}(m_0)} = \frac{b}{b - 1 + m_0^2(\beta)}.
		\label{eq:rate}
	\end{equation}
	
\begin{figure}[htp!]
		\center{\includegraphics[width=\linewidth]{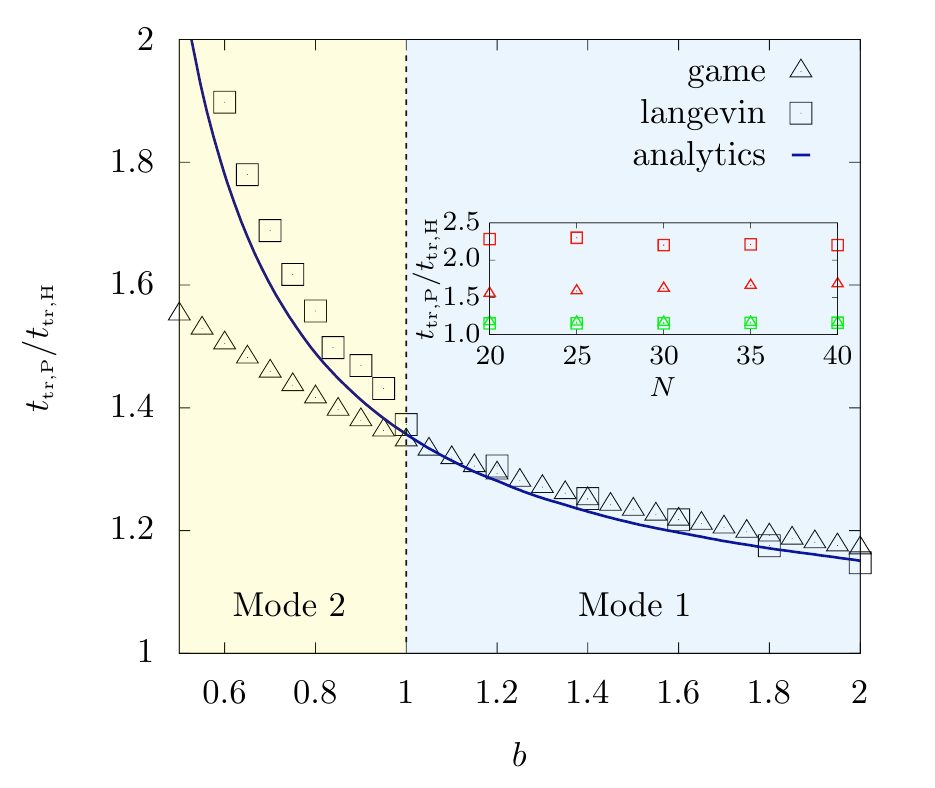}}
		\caption{Ratio of transition times in Hawkes and Poisson cases for $N = 20$. Triangles show simulation results for games (discrete model), and squares show results for Langevin equations (continuum model). The line refers to the theoretical prediction given by Eq.~\eqref{eq:rate}. The dashed line $b=1$ separates Mode 1 (blue area) from Mode 2 (yellow area). In the inset, we demonstrate the ratio of transition times as a function of agent number $N$ for $b = 0.5$ (red points) and $b = 2$ (green points) in both Langevin equation and Ising game numerical simulations (points are as in the legend for the main figure, and green points of both types are located lower than the corresponding red points). It is important to note that although the effect of a finite number of agents is present for both the Hawkes and Poisson cases, the impact of this effect is reduced when considering the ratio of transition times. Consequently, the overall dependence on the number of agents is weak.}
		\label{fig:transition}
	\end{figure}

	To check the above-formulated assumption we have performed computer simulations of Hawkes and Poisson Ising games as well as those of Langevin equations that correspond to Eq.~\eqref{eq:FPE}. A comparison of the results of these simulations with Eq.~\eqref{eq:rate} is shown in Fig.~\ref{fig:transition}. In its inset, we show that the transition time ratio weakly depends on the number of agents for $N \ge 20$ in two representative examples in Mode 1 and Mode 2 ($b = 2$, green points and $b = 0.5$, red points, respectively; green points of both types are located lower than the corresponding red points).
	
\begin{figure}[htp!]
		\center{\includegraphics[width=\linewidth]{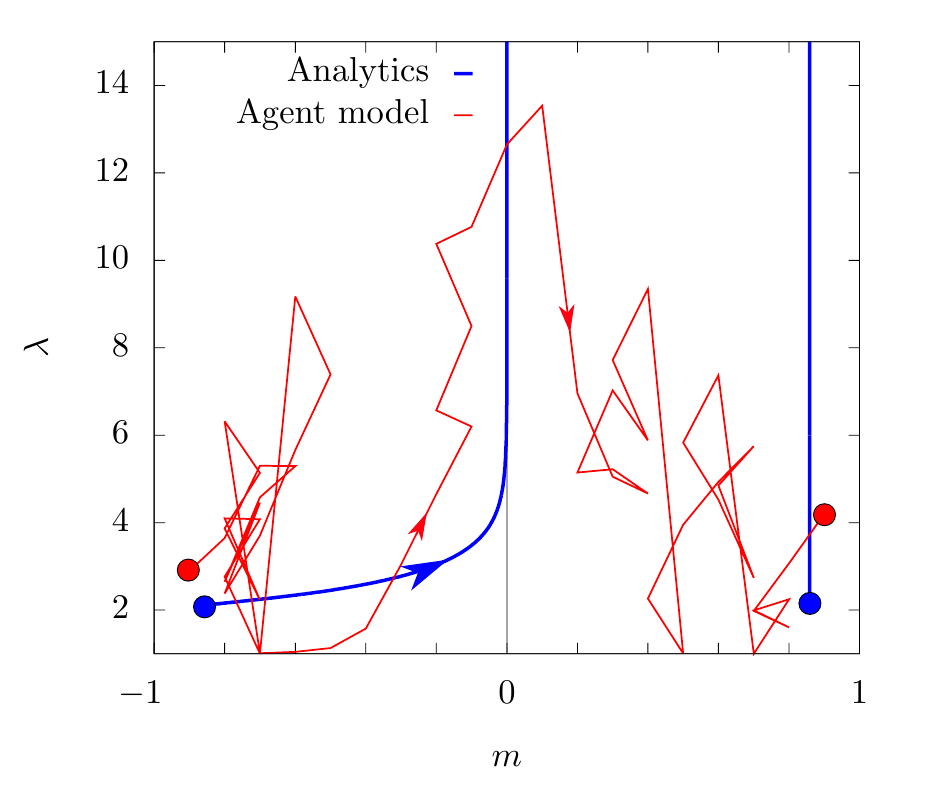}}
		\caption{Example of transition in Hawkes Ising game (red polyline) for $b = 0.5$, and $\lambda_0 = 1, N = 20$. As in the corresponding transition trajectory (blue curve), the intensity of decision making process increases near $m=0$.}
		\label{fig:mode_2}
	\end{figure}

	From Fig.~\ref{fig:transition} we see that the activity in the Hawkes Ising game as compared to the Poisson Ising game is indeed enhanced.  A more detailed conclusion is that in the regime corresponding to Mode 1 the formula in Eq.~\eqref{eq:rate} works well for the Mode 1 for both continuum and discrete cases, but in the regime corresponding to Mode 2 it is, due to the presence of divergence the continuum generalization of the game, not in agreement with the exact discrete formulation. Despite this, the shape of the transition trajectory still provides us a qualitatively correct insight into the behavior of agents, see Fig.~\ref{fig:mode_2}.
	
	Let us note that the decision process does significantly intensify around the separatrix, i.e.\ when are uncertain of which of the two (quasi)stable equilibria to choose. Once the decision is made, the agents calm down.
	
	\section{Conclusions}
	
	We have studied the self-excited Ising game on a complete graph. In spite of its simplicity, it has rich dynamics exhibiting various types of behavior. Competition of ``calming down'' and ``activation'' in the Hawkes self-excitation mechanism at different levels of noise results in three possible modes (phases). We expect that this competition might play an important role in other situations, e.g.,\ for non-exponential Hawkes kernels \cite{Bouchaud2018}, for more complicated graph topology, or for other types of noise in the agent utilities. The consideration of these situations also holds practical significance, as they can potentially exhibit a closer resemblance to real-world systems.
	
	Another focus in this work was to investigate the probability of transition between metastable equilibria in the infinite time limit. This is a very challenging task for a multi-dimensional case when the external field is non-gradient and has a discontinuity. Also, since in the relevant one-dimensional case (i.e.,\ when the potential field only has the discontinuity) it is known that the dynamics for such fields is rather different that for smooth potential fields \cite{Dekker1986}, it would be natural to assume a similar situation in the multi-dimensional case. However, based on the intuitive understanding of the considered model, we have presented an approach that allows us to reduce the problem to calculating the transition time in the corresponding one-dimensional model. This approach has also been validated by the numerical simulations. The analytically calculated transition trajectory also gave us a qualitative insight into the behavior of agents in the corresponding discrete system.
	
	As for further developments of the suggested approach, an interesting idea would be working out its generalization for two- and multi-dimensional systems. Compared to other another existing approaches for treating the case of non-gradient external field (see e.g.\ \cite{Cameron2021, Ashwin2022}), this method could present a workable alternative due to its simplicity.

\end{document}